          [2001/04/25 1.1 (PWD)]
\documentclass[a4paper,twocolumn]{esapub} 

\usepackage{times}
\usepackage{graphicx}

\title{HIGH ENERGY EMISSION FROM IGR J16320-4751}

\author[1]{L. Foschini}
\author[2]{J. A. Tomsick}
\author[3,4]{J. Rodriguez}
\author[4]{R. Walter}
\author[3]{A. Goldwurm} 
\author[3,5]{S.  Corbel}
\author[6]{P. Kaaret}

\affil[1]{IASF-CNR, Sezione di Bologna, Via Gobetti 101, 40129 Bologna (Italy)}
\affil[2]{Center for Astrophysics and Space Sciences, University of California at San Diego, La Jolla, CA 92093, USA}
\affil[3]{CEA Saclay, DSM/DAPNIA/SAp (CNRS FRE 2591), 91191, Gif--sur--Yvette Cedex, France}
\affil[4]{INTEGRAL Science Data Centre, Chemin d'Ecogia 16, 1290 Versoix, Switzerland}
\affil[5]{Universit\'e de Paris VII, F\'ed\'eration APC, 2 place Jussieu, 75251 Paris Cedex 05, France}
\affil[6]{Harvard--Smithsonian Center for Astrophysics, 60 Garden Street, Cambridge, MA 02138, USA}

\begin{document}


\maketitle

\begin{abstract}
IGR J$16320-4751$ was re--discovered by IBIS/ISGRI on board \emph{INTEGRAL} in early 
February 2003 during the observation of the black hole candidate 4U$1630-47$ (PI Tomsick). 
This source, already observed by ASCA and BeppoSAX, belongs to the class of heavily absorbed objects 
($N_H > 10^{23}$~cm$^{-2}$) that populate some arms of the Galaxy. Soon after the rediscovery 
by IBIS/ISGRI, the source was observed by \emph{XMM-Newton}: the arcsec position found with XMM 
allowed ones to find the most likely infrared counterpart. We present here the reanalysis of 
the high energy emission from IGR J$16320-4751$ detected by IBIS/ISGRI, including the spectral
and temporal characteristics. We also present a reanalysis of the \emph{XMM-Newton} and
optical/IR data.
\end{abstract}

\section{Introduction}
IGR J$16320-4751$ was serendipitously discovered on February 1.4, 2003 UT, 
with the IBIS/ISGRI detector (Ubertini et al. 2003, Lebrun et al. 2003) on board 
the \emph{INTEGRAL} satellite during the AO1 observation of the black hole candidate 
4U $1630-47$ (PI Tomsick). The coordinates (J2000) were $\alpha=16^{\rm h}:32^{\rm m}{\!}.0$ 
and $\delta=-47^{\circ}:51^{\rm m}{\!}.0$, with an uncertainty of $2'$ (Tomsick et al. 2003a). A quick 
archival research allowed ones to find immediately the X--ray counterpart in the ASCA 
catalog, as the source AX J$1631.9-4752$ (Sugizaki et al. 2001). IGR J$16320-4751$ is
therefore apparently located in the Norma Arm of the Galaxy, an
active star forming region distant from the Sun about $5$ kpc (Georgelin \& Georgelin 1976, 
Russeil 2003).

A Target of Opportunity request (ToO) with \emph{XMM-Newton} was immediately 
activated and it found one single source inside the error circle of ISGRI. 
The coordinates (J2000) were $\alpha=16^{\rm h}:32^{\rm m}:01^{\rm s}{\!}.9$ and 
$\delta=-47^{\circ}:52^{\rm m}:29^{\rm s}$ with an uncertainty of $4''$ (Rodriguez et al. 2003a). 
In this reduced error circle it was possible to identify two reliable infrared counterparts 
in the 2MASS catalog (Tomsick et al. 2003b).

A preliminary analysis of the \emph{XMM-Newton} observation, together with the identification
of the infrared counterpart was presented in Rodriguez et al. (2003b). Here we present 
a reanalysis of the \emph{INTEGRAL} AO1 observation, together with a reanalysis
of the \emph{XMM-Newton} ToO data.

\begin{figure*}
\centering
\includegraphics[scale=0.5]{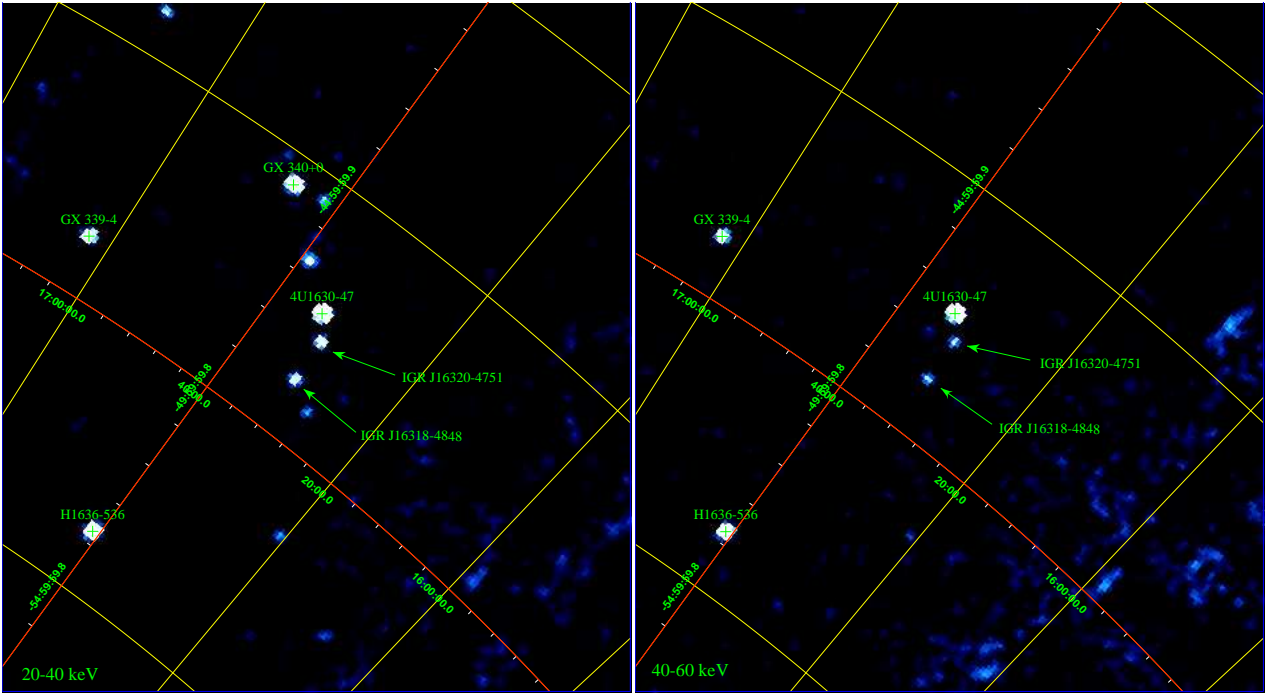}
\caption{IBIS/ISGRI mosaic of the Norma region centered on the BH 4U$1630-47$. 
(\emph{left}) Energy band $20-40$ keV, (\emph{right}) Energy band $40-60$ keV,
with a total exposure of $297$ ks. Equatorial coordinates are superimposed.
\label{fig:isgri_mosaic}}
\end{figure*}

\section{$INTEGRAL$ data analysis}
\emph{INTEGRAL} observed the Norma region from 1 February 2003 $05^{\rm h}:40^{\rm m}:58^{\rm s}$ to
5 February 2003 $07^{\rm h}:53^{\rm m}:01^{\rm s}$ UT for an elapsed time of $300$ ks and with
the dither pattern $5\times 5$. The screening, reduction, and analysis of the IBIS/ISGRI
data have been performed by using the \emph{INTEGRAL} Offline Scientific Analysis (OSA) 
v. 3 (Goldwurm et al. 2003a), available to the public through the \emph{INTEGRAL} Science 
Data Centre\footnote{\texttt{http://isdc.unige.ch/index.cgi?Soft+download}} (ISDC). 

\subsection{Imaging}
An initial analysis was performed by deconvolving the ISGRI shadowgrams
to obtain images. IGR J$16320-4751$ was detected in the imaging pipeline in 
the energy bands $20-40$ keV (signal--to--noise ratio $SNR=30\sigma$) and $40-60$ keV ($SNR=11\sigma$).
The mosaic images are shown in Fig. \ref{fig:isgri_mosaic}. Given the present
uncertainties in the validation of the software for the off--axis sources, it was
adopted the same procedure described by Goldoni et al. (2003). We divided the count
rate of IGR J$16320-4751$ by the count rate of the Crab in a similar off--axis angle, 
extracted from the calibration observations. We added a $5\%$ of systematic error, to
take into account the residual fluctuations in the count rate (cf Goldwurm et al. 2003b).
The calculated flux is $(8.0\pm 0.5)\times 10^{-11}$ erg cm$^{-2}$ s$^{-1}$ and
$(2.0\pm 0.2)\times 10^{-11}$ erg cm$^{-2}$ s$^{-1}$, in the energy bands $20-40$ keV and
$40-60$ keV respectively. No significant detection was recorded at higher energy.

\begin{figure*}
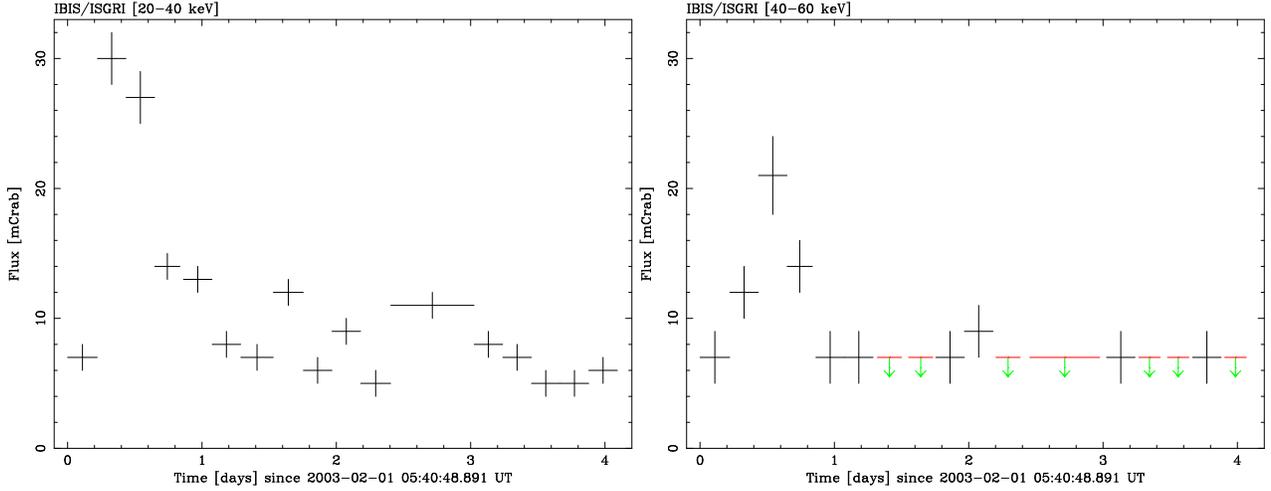

\centering
\includegraphics[angle=270,scale=0.36]{lc20_40.ps}
\includegraphics[angle=270,scale=0.36]{lc40_60.ps}
\caption{IBIS/ISGRI lightcurves of IGR J$16320-4751$ in the energy band $20-40$ keV (\emph{left}) and
$40-60$~keV (\emph{right}). Each bin is composed of 8 ScW, corresponding to about $15-20$ ks. 
Upper limits are at $3\sigma$ level.
\label{fig:isgri_lc}}
\end{figure*}

\subsection{Timing}
IGR J$16320-4751$ was detected in most of the individual ScWs with $3<SNR<6\sigma$. 
Therefore, to study the variability of the source it was decided to rebin the lightcurve 
so that each time bin contains 8 ScW, corresponding approximately to $15-20$ 
ks. The obtained lightcurve shows a clear outburst at the time of the discovery, plus some 
other periods with a certain variability 
(Fig. \ref{fig:isgri_lc}). During the outburst, the flux in the band $20-40$ keV raised
from about $7-8$ mCrab to $30$ mCrab (Fig.~\ref{fig:isgri_lc}, \emph{left}). The time scale of the 
variations occurs on $\approx 10^4$~s or even more. A similar behaviour is clearly seen in the 
$40-60$~keV energy band (Fig.~\ref{fig:isgri_lc}, \emph{right}).

Since in a coded--mask instrument the brightest sources in the field of view (FOV)
can -- under certain conditions -- significantly affect the detection of the other nearby
sources, we investigated also the time variability of the three brightest sources in the FOV,
namely 4U$1630-47$, 4U$1700-377$, and GX$340+0$. All the analysed sources display
different variability patterns and therefore we conclude that the variability of IGR 
J$16320-4751$ is genuine.

\subsection{Spectral extraction}
To extract the spectrum of the whole observation and also during the outburst we used
both the spectral extraction pipeline and the results from the image analysis, to perform
a check of the results. This is necessary to take into account that IGR J$16320-4751$ is 
located in a crowded region, with at least two nearby sources (4U $1630-47$ and IGR J$16318-4848$),
and therefore the different approach in the deconvolution of images and extraction of spectra
(see Goldwurm et al. 2003a for a full explanation of the algorithms) could generate results not 
always consistent each others.

For the spectral extraction pipeline we rebinned the latest RMF matrix from the original $2048$ channels 
to $21$ channels, by putting all the channels above $200$ keV into the last channel of the rebinned matrix. 
The remaining channels in the range $15-200$ keV were grouped in bin of about $10$ keV in size. This approach 
has been selected to emphasize the energy range $15-200$ keV and to have enough statistics to perform 
the $\chi^2$ test in the spectral fit with \texttt{xspec} (v 11.3). Nonetheless, the source is faint and does
not allow a fit with multiple component models. We decided to fit with a simple
power law model to measure if there is hardening/softening during the outburst.

The spectrum of the whole observation has $\Gamma=3.8\pm 0.9$, with a flux of
about $9\times 10^{-11}$ erg cm$^{-2}$ s$^{-1}$ in the energy band $20-60$ keV, consistent
with the results from the imaging analysis. Also in the spectral extraction
pipeline no flux is detected for energies greater than $\approx 60$ keV.

The spectrum extracted during of the outburst (bins 2-4 in Fig. \ref{fig:isgri_lc})
has the photon index varying from $\Gamma\approx 3.1$ to $\Gamma\approx 2.6$, therefore slightly harder than
that of the whole observation, with a peak flux of $2\times 10^{-10}$ erg cm$^{-2}$ s$^{-1}$
in the $20-60$ keV energy band.

During the ``normal'' activity, investigated only with the imaging pipeline, IGR J$16320-4751$ is 
barely detected in the energy band $40-60$ keV (Fig.~\ref{fig:isgri_lc}, \emph{right}), with the
exception of the outburst.

The hardness ratio (not shown), calculated by using the fluxes in the two energy bands 
($20-40$~keV and $40-60$~keV),
is in agreement with the results of the spectral extraction pipeline, and
confirms the softness of the source during almost the whole observation.

\begin{figure}
\centering
\includegraphics[scale=0.5]{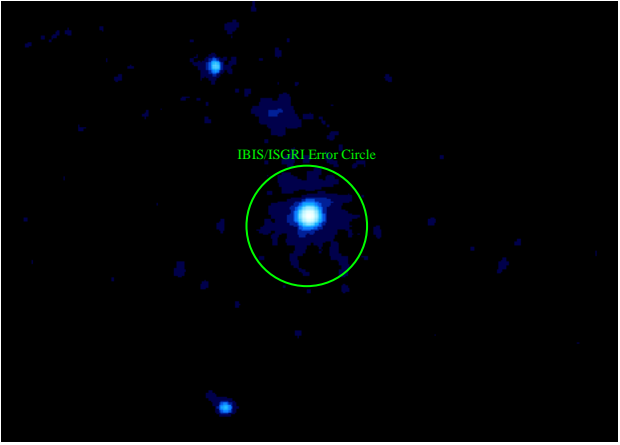}
\caption{EPIC--PN image in the $0.2-12$ keV energy band, with $5$ ks of exposure (i.e. cleaned
from soft-proton flares). The IBIS/ISGRI error circle is superimposed. It is clearly visible 
only one source inside the error circle.
\label{fig:epic_pn}}
\end{figure}

\section{$XMM-Newton$ data analysis}

\begin{figure*}
\centering
\includegraphics[angle=270,scale=0.6]{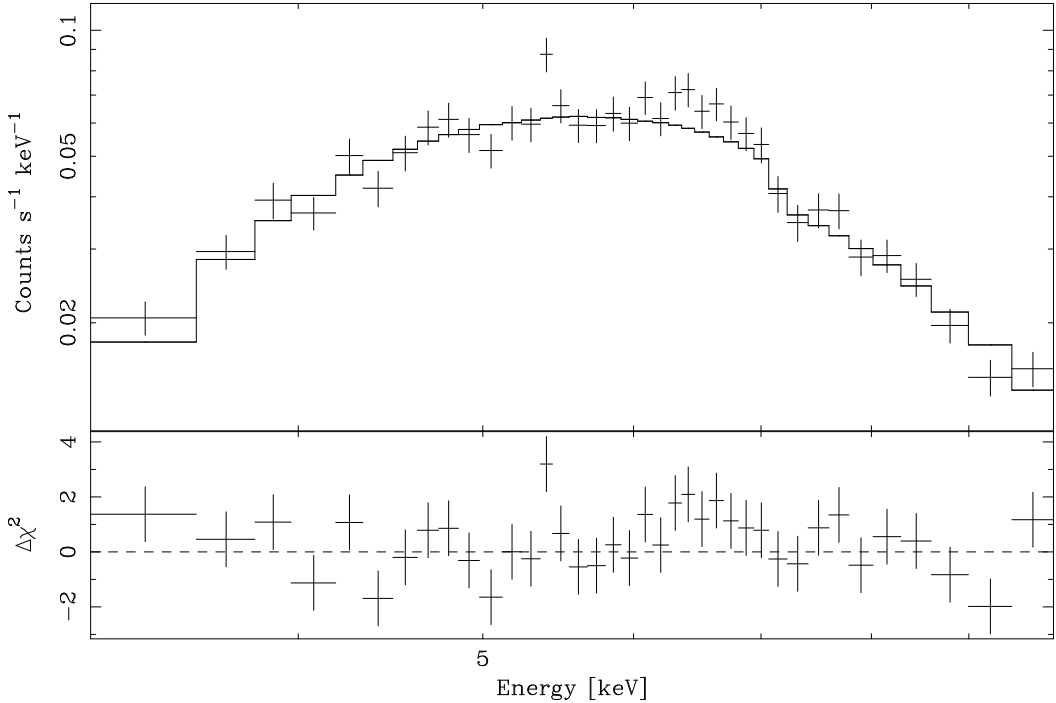}
\caption{EPIC--PN spectrum in the $3-12$ keV energy band, with $17$ 
ks of exposure and fitted with a single power law model. 
The deviations of data versus model are shown in the bottom window. 
\label{fig:epic_spec}}
\end{figure*}

\emph{XMM-Newton} observed IGR J$16320-4751$ as a Target of Opportunity (ToO) from 4 March 2003
$20^{\rm h}:20^{\rm m}:38^{\rm s}$ UT to 5 March 2003 $03^{\rm h}:47^{\rm m}:18^{\rm s}$ UT, with an 
elapsed time of about $26$ ks.
Due to soft-proton flares, the effective exposure was about $5$ ks, but sufficient
to clearly identify (uncertainty $<4''$) the X--ray counterpart at coordinates (J2000) 
$\alpha=16^{\rm h}:32^{\rm m}:01^{\rm s}{\!}.9$ and 
$\delta=-47^{\circ}:52^{\rm m}:29^{\rm s}$ (Rodriguez et al., 2003a,b; see Fig. \ref{fig:epic_pn}). An early 
analysis of the \emph{XMM-Newton} data (ToO) were published in Rodriguez et al. (2003b). The lightcurve
in the $5$ ks clean from soft--proton flares showed a time variability on time scales of $100$ s,
while the spectrum was found to be harder than that of ISGRI, fitted with an absorbed power
law with photon index $\Gamma=1.6_{-0.1}^{+0.2}$ and $N_{\rm H}=(2.1_{-0.1}^{+0.4})\times 10^{23}$ cm$^{-2}$
($\chi^{2}=55.9$, d.o.f.$=69$). The unabsorbed flux in the $2-10$ keV energy band was $1.7\times 10^{-11}$
erg cm$^{-2}$ s$^{-1}$. We refer the reader to the paper by Rodriguez et al. (2003b) for more details.

For the present work, we decided to reanalyze the \emph{XMM-Newton} data to extract 
the better spectrum. We used still the XMM SAS v. 5.4.1 software to process and screening the data and
the same procedures described in Rodriguez et al. (2003b). We adopted a different procedure only
to extract the spectrum. Since the source is bright enough to be clearly visible still in the flared image,
we avoided the selection of a time region not affected by soft-proton flares, and we extracted directly
from the contaminated image the source plus background counts from a circular region centered on 
IGR~J$16320-4751$ with radius of $30''$. Therefore, we extracted the background from another circular
region with radius $2'$, and close to the source. We performed the background (now including the
soft--proton flares) subtraction directly in \texttt{xspec}. 

We succeeded to save $17$ ks from the elapsed time of $26$ ks, with a loss of only $9$ ks (to be 
compared with the loss of $21$ ks of the early analysis). We found that the best fit model is still 
an absorbed power law with $\Gamma=1.4_{-0.1}^{+0.2}$ and $N_{\rm H}=(2.2\pm 0.2)\times 10^{23}$ cm$^{-2}$
($\chi^{2}=204.5$, d.o.f.$=184$), consistent with the results of the early analysis 
(see Fig. \ref{fig:epic_spec}). The unabsorbed flux in the $0.2-12$ keV energy band is 
$2.8\times 10^{-11}$ erg cm$^{-2}$ s$^{-1}$.

We note that there are some interesting features between $4$ and $10$ keV, and particularly some
hints for emission lines of the iron complex, between $6$ and $7$ keV. The addition of a thermal
plasma model (\texttt{mekal} model in \texttt{xspec}) with a temperature $kT=5.5$ keV is able to reproduce
some of these features, and particularly the emission lines of the iron complex. However, this model is
statistically required only at $91\%$ confidence level. 
A simple gaussian emission line at $E=6.5\pm 0.1$~keV and width $\sigma=0.3_{-0.1}^{+0.2}$~keV 
is required at $99.67$\%. 
Further observations are required to better constrain the nature of this excess.

\begin{table}[!ht]
  \begin{center}
    \caption{Optical/Infrared counterparts of IGR J$16320-4751$.}\vspace{1em}
    \renewcommand{\arraystretch}{1.2}
    \begin{tabular}[h]{ll}
	\hline
	\textbf{Source 1} & {}\\
      \hline
      Catalog & Magnitude \\
      \hline
	2MASS$^{\mathrm{a}}$   & $J<14.08$\\
	2MASS$^{\mathrm{a}}$   & $H=13.03\pm 0.04$\\
	2MASS$^{\mathrm{a}}$   & $K=10.99\pm 0.04$\\ 
	\hline
	\textbf{Source 2} & {}\\
	\hline
	USNO B1$^{\mathrm{b}}$ & $B_1$ n.a.\\
	USNO B1$^{\mathrm{b}}$ & $B_2=17.3\pm 0.3$\\
	USNO A2$^{\mathrm{b}}$ & $B=17.3\pm 0.3$\\
	GSC 2.2$^{\mathrm{c}}$  & $B=18.0\pm 0.4$\\
	USNO CCD AC$^{\mathrm{d}}$ & $UCAC=16.0\pm 0.3$\\
	USNO B1$^{\mathrm{b}}$ & $R_1=14.6\pm 0.3$\\
	USNO B1$^{\mathrm{b}}$ & $R_2=15.4\pm 0.3$\\
	USNO A2$^{\mathrm{b}}$ & $R=15.0\pm 0.3$\\
	GSC 2.2$^{\mathrm{c}}$  & $R=15.4\pm 0.4$\\
	USNO B1$^{\mathrm{b}}$ & $I=14.2\pm 0.3$\\
	DENIS$^{\mathrm{e}}$   & $I=14.00\pm 0.03$\\
	2MASS$^{\mathrm{a}}$   & $J=12.13\pm 0.02$\\
	DENIS$^{\mathrm{e}}$   & $J=12.22\pm 0.09$\\
	2MASS$^{\mathrm{a}}$   & $H=11.24\pm 0.03$ \\
	2MASS$^{\mathrm{a}}$   & $K=10.82\pm 0.04$ \\
	DENIS$^{\mathrm{e}}$   & $K=10.75\pm 0.07$\\
      \hline \\
      \end{tabular}
    \label{tab:OIR}
  \end{center}
\begin{list}{}{}
\item[$^{\mathrm{a}}$] Two Microns All Sky Survey (2MASS) Point Source catalog, Cutri et al. (2003).
\item[$^{\mathrm{b}}$] US Naval Observatory Catalog A2, Monet et al. (1998). US Naval Observatory Catalog
B1, Monet et al. (2003). For the latter, B and R magnitudes are taken from two types of plates and referenced
with subscripts 1 and 2: (1) Palomar Observatory Sky Survey (POSS) I, ($1949-1965$), with emulsion sensible at 
the wavelengths $620-670$ nm; (2) POSS II ($1985-2000$), sensible at $385-540$ nm.
\item[$^{\mathrm{c}}$] Guide Star Catalog 2.2.
\item[$^{\mathrm{d}}$] US Naval Observatory CCD Astrograph Catalog (UCAC, \texttt{http://ad.usno.navy.mil/ucac/}). 
UCAC magnitude is in the wavelenght band $579-642$ nm, between V and R bands.
\item[$^{\mathrm{e}}$] DEep Near Infrared Survey (DENIS) of the southern sky (DENIS Consortium 
2003, \texttt{http://cdsweb.u-strasbg.fr/denis.html}). 
\end{list}
\end{table}

\begin{figure*}
\centering
\includegraphics[scale=0.33]{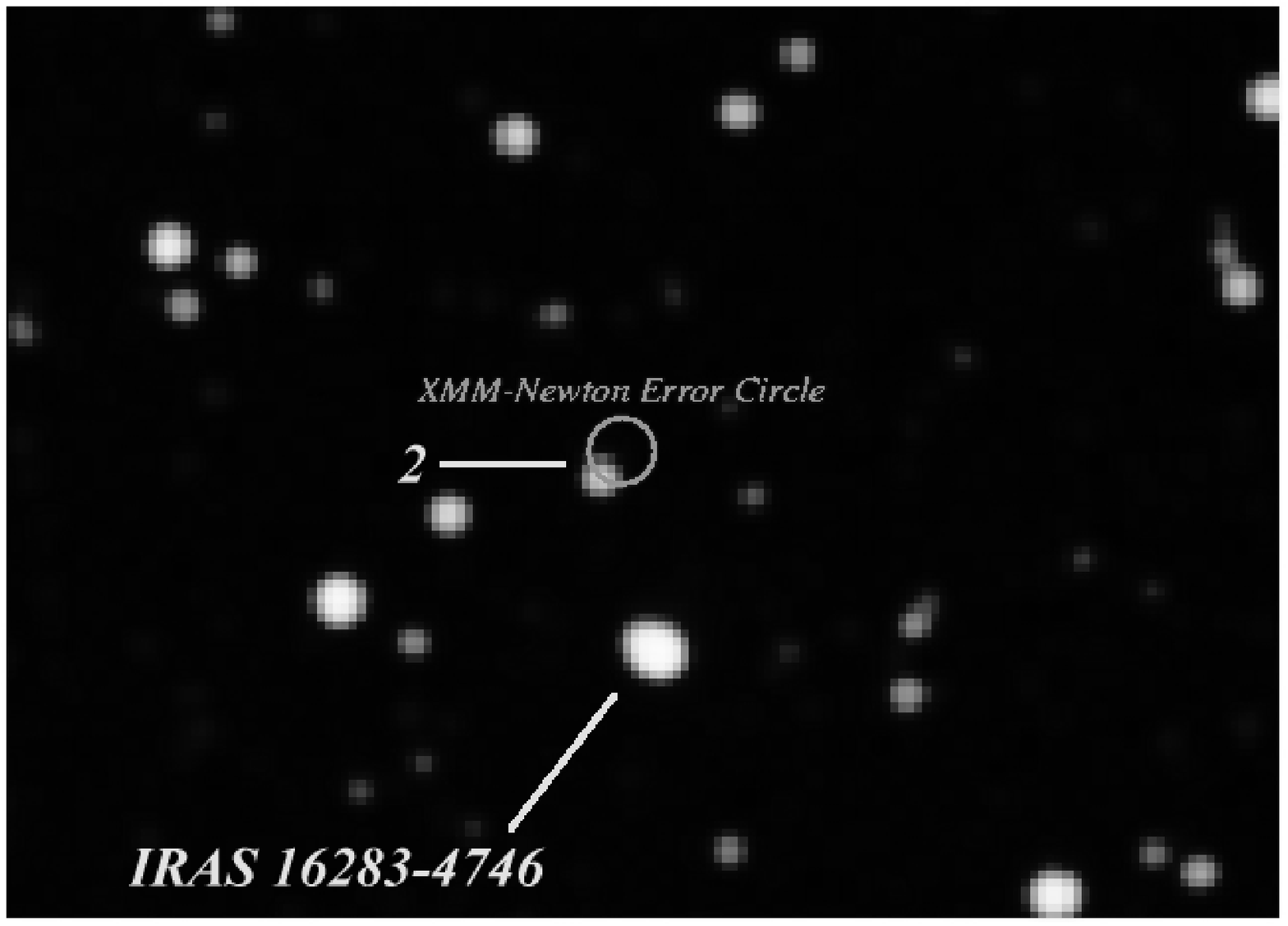}
\includegraphics[scale=0.33]{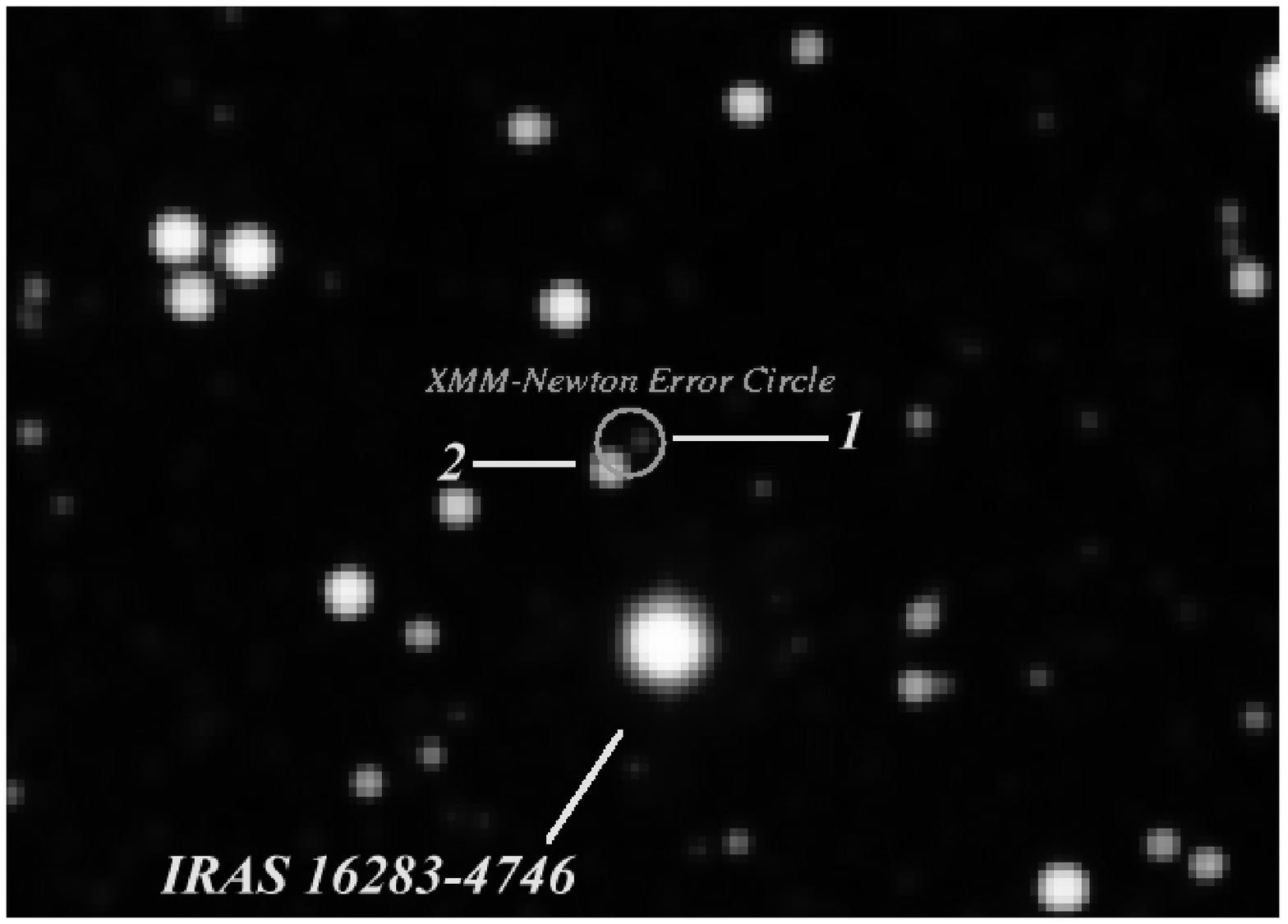}
\includegraphics[scale=0.33]{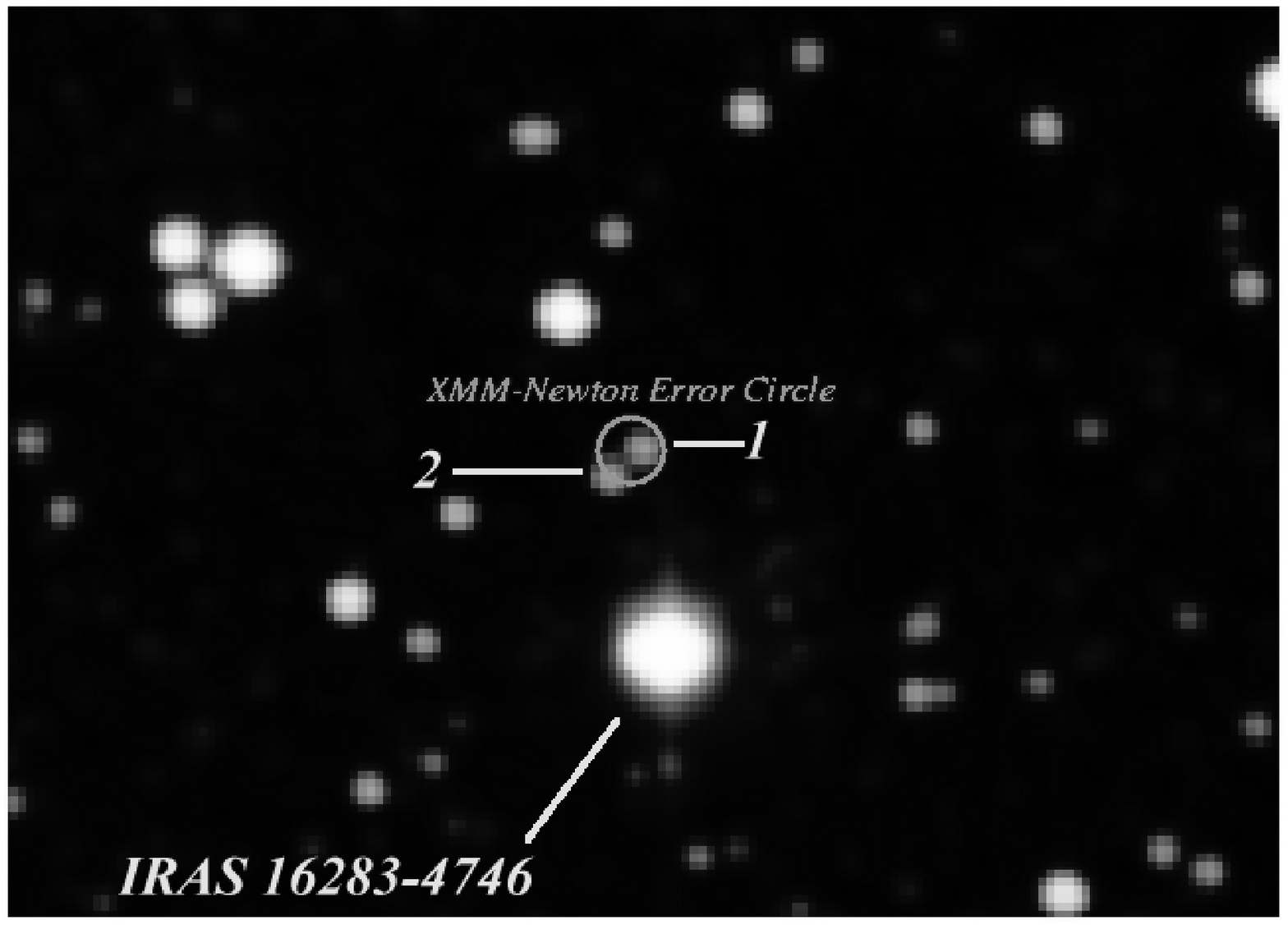}
\caption{2MASS images in the bands J (\emph{left}), H (\emph{centre}), and K (\emph{right}). 
The \emph{XMM-Newton} error circle is superimposed.
\label{fig:2MASS}}
\end{figure*}

\section{Optical/Infrared Counterparts}
With the improved position given by \emph{XMM-Newton}, it became possible to look for the counterparts
at other wavelengths, specifically in the optical/infrared bands (Tomsick et al. 2003b). Two sources 
have been found in the Two Microns All--Sky Survey (2MASS): the first, labelled 1 in Fig.~\ref{fig:2MASS},
could also be the most probable counterpart. The infrared colours along the line of sight ($E(J-H)=1.0$, $E(H-K)=0.8$, 
for the extinction due to the Galactic absorption $A_V=11.1$) suggesting an excess perhaps due to the 
presence of hot plasma or circumstellar dust, consistent with the findings of the X--ray analysis. 
Specifically, the source 1 is not detected in the $J$ band, with an upper limit of $J>14$. Concerning the source 2, there are detections
in other catalogs: a summary of the optical/infrared detections is shown in Table~\ref{tab:OIR}. It is
worth noting that the source 2 is classified as a non-star object, that may be either galaxies or blended objects,
in the Guide Star Catalog 2.2{\footnote{\texttt{http://www-gsss.stsci.edu/gsc/GSChome.htm}}}. 
Indeed, it appears slightly elongated with an eccentricity of $0.07$ and a semimajor axis of $3.12$ pixels.

\begin{figure}
\centering
\includegraphics[angle=270,scale=0.35]{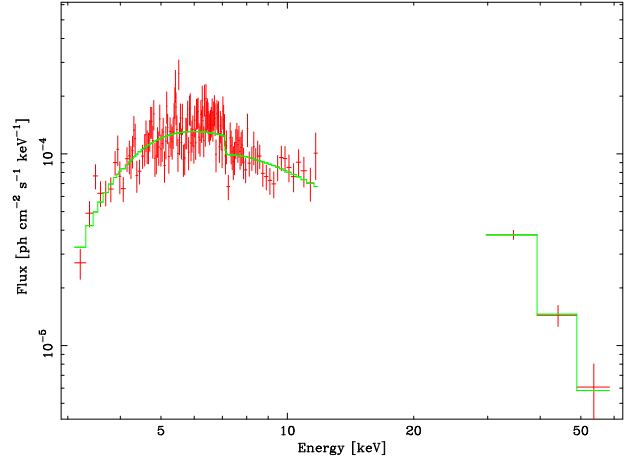}
\caption{Broad band unfolded spectrum of IGR~J$16320-4751$  obtained with a simultaneous fit
of \emph{XMM-Newton} (EPIC--PN) and \emph{INTEGRAL} (IBIS/ISGRI) data. The
energy ranges are $3-12$~keV for \emph{XMM-Newton} and $30-60$~keV for \emph{INTEGRAL}.
\label{fig:allspec}}
\end{figure}

\section{Discussion and Conclusions}
In the present work we confirmed and extended the results obtained by Rodriguez et al. (2003b).
The nature of IGR~J$16320-4751$ is not clear yet and, even though the most probable hypothesis is
a Galactic X--ray binary, the possibility of an extragalactic object cannot be completely ruled out.
The source appears to be intrinsically absorbed, with an $N_{H}$ one order of magnitude greater than
the Galactic absorption along the line of sight. Also the infrared observation suggests
an excess due to circumstellar plasma or dust along the line of sight. The X--ray spectrum could support the
hot plasma solution, but the confidence level is not enough (only $91\%$) and should be confirmed with
further observations with much more statistics. Therefore, the picture of a high--mass X--ray binary 
(HMXRB) with a compact object accreting through winds appears to be very likely for IGR~J$16320-4751$. 
The winds could also be the source of obscuration, wrapping up the compact object. 

The nature of the accreting object is not clear as well: the strong differences in the photon index
in the energy bands $0.5-10$~keV and $20-60$~keV could be due to changes in spectral states, thus
suggesting a black hole. On the other hand, the drop of emission for energies greater than $60$~keV
is common in neutron stars. Although, the \emph{INTEGRAL} and \emph{XMM-Newton} are not simultaneous,
we tried a simultaneous fit (Fig.~\ref{fig:allspec}). The best fit model ($\chi^2=344.8$, $dof=344$) is an 
absorbed power law with exponential cutoff (\texttt{cutoffpl} model in \texttt{xspec}), typical 
of accreting neutron stars (see, e.g. White et al. 1983). The photon index is $\Gamma=0.7_{-0.3}^{+0.2}$, 
the column density $N_H=(2.0\pm 0.2)\times 10^{23}$~cm$^{-2}$, and the cutoff energy $10\pm 3$~keV. The 
scaling constant applied to the ISGRI spectrum is $5_{-2}^{+3}$. We
stress that the observations were \emph{not} simultaneous and the source is strongly variable.
Therefore, this fit should be taken with extreme care.

Another possibility, given the low luminosity in both bands ($8\times 10^{34}$~erg/s
in the $0.2-12$~keV energy band, and $3\times 10^{35}$~erg/s in the $20-60$~keV band) if the source is 
located in the Norma Arm ($5$~kpc), is that we are observing the emission from a jet. In this case, we could make 
the hypothesis that the change in the photon index could be due to a break, because of a change in
the mechanism of cooling of electrons.

The key question in the study of this source is therefore the spectral variability. An approved 
coordinated \emph{INTEGRAL} and \emph{XMM-Newton} observation to be performed by next August
should give us sufficient data to disentangle these hypotheses.

\section*{Acknowledgments}
Based on observations obtained with \emph{INTEGRAL}, an ESA mission with instruments and science
data centre funded by ESA Member States (especially the PI countries:
Denmark, France, Germany, Italy, Switzerland, Spain), Czech Republic and
Poland, and with the participation of Russia and the USA.
Based on observations obtained with \emph{XMM-Newton}, an ESA science mission
with instruments and contributions directly funded by ESA Member States
and the USA (NASA).
This research has made use of data obtained from the High Energy Astrophysics 
Science Archive Research Center (HEASARC), provided by NASA's Goddard Space Flight Center,
and of data products from the Two Micron All Sky Survey, 
which is a joint project of the University of Massachusetts and the Infrared Processing 
and Analysis Center/California Institute of Technology, funded by the National 
Aeronautics and Space Administration and the National Science Foundation.
    
LF acknowledges the Italian Space Agency (ASI) for financial support.


\begin{thebibliography}{}

\bibitem[]{cameron} Cameron L.M., 1990, A\&A 233, 16
\bibitem[]{cutri} Cutri R.M., Skrutskie M.F., van Dyk S., et al., 2003, 2MASS All--Sky Catalog of Point Sources.
University of Massachusetts and Infrared Processing and Analysis Center, 
(IPAC/California Institute of Technology)

\bibitem[]{georg} Georgelin Y.M., Georgelin Y.P., 1976, A\&A 49, 57

\bibitem[]{paolo} Goldoni, P., Bonnet--Bidaud, J.M., Falanga, M., \& Goldwurm, A., 2003, A\&A 411, L399

\bibitem[]{goldwurm1} Goldwurm A., David P., Foschini L., et al., 2003a, A\&A 411, L223
\bibitem[]{goldwurm2} Goldwurm, A., Gros, A., Goldoni, P., et al., 2003b, IBIS/ISGRI
Instrument Specific Software Scientific Validation Report, v. 1.0 



\bibitem[]{lebrun} Lebrun F., Leray J.-P., Lavocat P., et al., 2003, A\&A 411, L141

\bibitem[]{monet1} Monet D.G., Bird A., Canzian B., et al., 1998, U.S. Naval Observatory Flagstaff Station 
(USNOFS) and Universities Space Research Association (USRA) stationed at USNOFS.

\bibitem[]{monet2} Monet D.G., Levine S.E., Casian B., et al., 2003, AJ 125, 984

\bibitem[]{rodriguez1} Rodriguez J., Tomsick J.A., Foschini L., et al., 2003a, IAUC 8096
\bibitem[]{rodriguez2} Rodriguez J., Tomsick J.A., Foschini L., et al., 2003b, A\&A 407, L41


\bibitem[]{russeil} Russeil D., 2003, A\&A 397, 133

\bibitem[]{sugizaki} Sugizaki et al., 2001, ApJS 134, 77

\bibitem[]{tomsick1} Tomsick J.A., Lingenfelter R., Walter R., et al., 2003a, IAUC 8076
\bibitem[]{tomsick2} Tomsick J.A., Rodriguez J., Goldwurm A., et al., 2003b, IAUC 8096

\bibitem[]{uber} Ubertini P., Lebrun F., Di Cocco G., et al., 2003, A\&A 411, L131

\bibitem[]{white} White N.E., Swank J.H., Holt S.S., 1983, ApJ 270, 711


\end{thebibliography}
\end{document}